\documentclass[11pt,preprint]{aastex}

\begin{document}

\title{High energy cosmic-rays from gamma-ray burst sources: A stronger case}

\author{Eli Waxman\altaffilmark{1}}
\altaffiltext{1}{Physics Faculty, Weizmann Institute, Rehovot
76100, Israel; waxman@wicc.weizmann.ac.il}

\begin{abstract}

The suggested association between the sources of $\gamma$-ray
bursts (GRB's) and the sources of ultra-high energy cosmic rays
(UHECR's) is based on two arguments: (i) The constraints that UHECR
sources must satisfy to allow proton acceleration to $>10^{20}$~eV
are similar to those inferred for GRB sources from $\gamma$-ray
observations, and (ii) The average energy
generation rate of UHECR's is similar to the $\gamma$-ray
generation rate of GRB's. We show that recent GRB and UHECR observations
strengthen both arguments, and hence strengthen the suggested
association.

\end{abstract}


\section{Introduction}
\label{sec:introduction}

The widely accepted interpretation of the phenomenology of
$\gamma$-ray bursts (GRB's), bursts of 0.1 MeV--1 MeV photons
lasting for a few seconds \citep[see][for review]{Fishman95}, is
that the observable effects are due to the dissipation of the
kinetic energy of a cosmologically distant, relativistically
expanding wind, a ``fireball,'' whose primal cause is not yet
known \citep[for reviews see][]{fireballs1,fireballs2,fireballs3}. 
\citet{W95a}, \citet{MnU95} and \citet{Vietri95} have suggested that ultra-high energy,
$>10^{19}$~eV, cosmic rays (UHECR's) may be produced in GRB sources.
The model suggested in \citep{W95a} was based on two
arguments. First, it was shown that the constraints imposed on the
relativistic wind by the requirement that it produces observed GRB
characteristics are similar to the constraints imposed on such a
wind by the requirement that it would allow proton acceleration to
$>10^{20}$~eV. Second, the energy generation rate of $\gamma$-rays
by GRB's was shown to be similar to the energy generation rate
required to account for the observed UHECR flux \citep{W95a,W95b}.

The origin of UHECR's is one of the most exciting open questions
of high energy astrophysics \citep{Sigl,NaganoWatson}. The extreme
energy of the highest energy events poses a challenge to models of
particle acceleration. Since very few known astrophysical objects
have characteristics indicating that they may allow acceleration
of particles to the observed high energies \citep{pascos03}, the question of
whether GRB's are possible UHECR sources is of great interest.
Moreover, since the GRB model for UHECR production makes unique
predictions, which differ from those of other models 
\citep[see][and discussion in \S~\ref{sec:discussion}]{W01rev}, the design and analysis of
future large area UHECR experiments may be affected by the answer
to this question.

The detection over the past few years of "afterglows," delayed
low energy (X-ray to radio) emission of GRBs \citep[for review see][]{AG_ex_review}
confirmed the cosmological origin
of the GRB's, through redshift determination of GRB host-galaxies,
and confirmed standard model predictions of afterglows that result
from the collision of an expanding fireball with its surrounding
medium \citep[e.g.][]{fireballs1,fireballs2,fireballs3}. In addition to
providing strong support to the fireball model, these observations
also provide new constraints on fireball model
parameters and  more accurate information on the redshift distribution of GRB sources.
Recently, new data on the spectrum and flux of UHECR's was
presented by the HiRes experiment \citep{HiRes}, providing improved
constraints on the generation rate and spectrum of UHECR's
\citep{BnW02}. Here we discuss the implications of these new GRB and UHECR
observations to the GRB model for UHECR production. 

In \S~\ref{sec:model_description} we briefly describe the model proposed in
\citep{W95a} for proton acceleration in GRB fireballs. The main goal of this section is to identify the key constraints that the relativistic wind parameters need to satisfy, Eqs.~(\ref{eq:xiB}) and (\ref{eq:Gmin}), in order to allow acceleration of protons to $>10^{20}$~eV \citep[A more detailed and pedagogical description of the model is given in][]{W01rev}. The association of GRB and UHECR sources was motivated mainly by the fact that these constraints were similar to those inferred, based on independent physical arguments, from $\gtrsim1$~MeV $\gamma$-ray observations. Recent lower energy afterglow observations provide new constraints on model parameters, which are independent of the $\gamma$-ray constraints. In \S~\ref{sec:model_observations} we show that the afterglow constraints are similar to the $\gamma$-ray constraints, and in fact imply parameter values which are more favorable for the acceleration of protons to $>10^{20}$~eV. In \S~\ref{sec:model_other} we compare our results to those of other authors. In particular, we show that results derived by several authors, arguing that the maximum proton energy is too low \citep{Gallant99,Achterberg01} or
that the proton spectrum is too steep \citep{Ostrowski02} to
account for the observed UHECR spectrum, are not applicable to the model proposed in \citep{W95a}. 

The determination of GRB redshifts, made possible by afterglow detection, lead to significant changes in, and to significantly reduced uncertainties of, the estimates of both GRB rate and average $\gamma$-ray energy release (per single GRB). We show in \S~\ref{sec:rate_GRB} that the local energy production rate in $\gamma$-rays by GRB's, inferred using recent redshift measurements, is similar to the pre-afterglow estimate. In \S~\ref{sec:rate_CR} we show that the local rate of energy production in UHECR's inferred using recent UHECR observations is consistent with, although more accurate than, earlier estimates \citep{W95b}, and hence comparable to the energy production rate in $\gamma$-rays by GRB's. In \S~\ref{sec:rate_other} we demonstrate that recent claims to the contrary \citep{Stecker00,Berezinsky02}, in particular that the UHECR energy production rate is 3 order of magnitude higher than the GRB $\gamma$-ray production rate~\citep{Berezinsky02}, are erroneous. 

Our main results are summarized in \S~\ref{sec:discussion}.

\section{Proton acceleration in GRB fireballs}
\label{sec:model}

\subsection{Brief description of the model}
\label{sec:model_description}

General phenomenological considerations, based on $\gamma$-ray
observations, indicate that, regardless of the nature of the
underlying sources, GRB's are produced by the dissipation of the
kinetic energy of a relativistic expanding fireball. A compact
source, $r_0\sim10^7$~cm, produces a wind, characterized by an
average luminosity $L\sim10^{52}{\rm erg\ s}^{-1}$ and mass loss
rate $\dot M$. At small radius, the wind bulk Lorentz factor,
$\Gamma$, grows linearly with radius, until most of the wind
energy is converted to kinetic energy and $\Gamma$ saturates at
$\Gamma\sim L/\dot M c^2\sim300$. Variability of the source on a
time scale $\Delta t\sim10$~ms, resulting in fluctuations in the
wind bulk Lorentz factor $\Gamma$ on a similar time scale, results
in internal shocks in the ejecta at a radius $r\sim
r_d\approx\Gamma^2c\Delta t\gg r_0$. It is assumed that internal
shocks reconvert a substantial part of the kinetic energy to
internal energy, which is then radiated as $\gamma$-rays by
synchrotron and inverse-Compton radiation of shock-accelerated
electrons. At a later stage, the shock wave driven into the
surrounding medium by the expanding fireball ejecta leads to the
emission of the lower-energy afterglow.

The observed radiation is produced, both during the GRB and the
afterglow, by synchrotron emission of shock accelerated electrons.
In the region where electrons are accelerated, protons are also
expected to be shock accelerated. This is similar to what is
thought to occur in supernovae remnant shocks, where synchrotron
radiation of accelerated electrons is the likely source of
non-thermal X-rays, and where shock acceleration of protons is
believed to produce cosmic rays with energy extending to
$\sim10^{15}{\rm eV}$ \citep[e.g.][]{SN1006,SN1006CR,Tycho}. 
Thus, it is likely that protons, as well as
electrons, are accelerated to high energy within GRB fireballs.

The internal shocks within the expanding wind are expected to be mildly
relativistic in the wind rest frame, due to the fact that the allowed range of Lorentz
factor fluctuations within the wind is from few~$\times10^2$ (the lower limit required to avoid large optical depth) to few~$\times10^3$ \citep[the maximum Lorentz factor to which shell acceleration by radiation pressure is possible, e.g.][]{fireballs3}. This implies that the Lorentz factors associated
with the relative velocities are not very large. Since internal shocks are mildly
relativistic, we expect our
understanding of non-relativistic shock acceleration to apply to
the acceleration of protons in these shocks. In particular, the
predicted energy distribution of accelerated protons is expected
to be $dn_p/dE_p\propto E_p^{-2}$ \citep{AXL77,Bell78,BnO78},
similar to the predicted electron energy spectrum, which is
consistent with the observed photon spectrum.

Several constraints must be satisfied by wind parameters in order
to allow proton acceleration to high energy $E_p$. We summarize
below these constraints. The reader is referred to
\citep{W95a,W01rev} for a detailed derivation. The requirement that
the acceleration time be smaller than the wind expansion time
(which also implies that the proton is confined to the
acceleration region over the required time) sets a lower limit to
the strength of the wind magnetic field. This may be expressed as
a lower limit to the ratio of magnetic field to electron energy
density \citep{W95a},
\begin{equation}
u_B/u_e>0.02 \Gamma_{2.5}^2 E_{p,20}^2L_{\gamma,52}^{-1},
\label{eq:xiB}
\end{equation}
where $E_p=10^{20}E_{p,20}$~eV, $\Gamma=10^{2.5}\Gamma_{2.5}$ and
$L_{\gamma}=10^{52}L_{\gamma,52}{\rm erg/s}$ is the wind
$\gamma$-ray luminosity. A second constraint is imposed by the
requirement that the proton acceleration time be smaller than the
proton energy loss time, which is  dominated by synchrotron
emission. This sets an upper limit to the magnetic field strength,
which in turn sets a lower limit to $\Gamma$ \citep{W95a,RnM98}
\begin{equation}
\Gamma>130 E_{p,20}^{3/4}\Delta t^{-1/4}_{-2}. \label{eq:Gmin}
\end{equation}
Here, $\Delta t=10^{-2}\Delta t_{-2}$~s. As explained in
\citep{W95a}, the constraints Eq.~(\ref{eq:xiB}) and Eq.~(\ref{eq:Gmin})
hold regardless of whether the fireball is a sphere or a narrow
jet (as long as the jet opening angle is $>1/\Gamma$). The
luminosity in Eq.~\ref{eq:xiB} is the "isotropic equivalent
luminosity", i.e. the luminosity under the assumption of isotropic
emission.

Internal shocks within the wind take place at a radius $r_d\approx\Gamma^2c\Delta t$. The constraint of Eq.~(\ref{eq:xiB}) is independent of $\Delta t$, i.e. independent of the internal collision radius, while the constraint of Eq.~(\ref{eq:Gmin}) sets a lower limit to the collision radius for a given $\Delta t$. This implies that protons may be accelerated to $>10^{20}$~eV regardless of the value of $\Delta t$, which may range from the dynamical time of the source ($\Delta t\sim1$~ms) to the wind duration ($\Delta t\sim1$~s), provided the magnetization and Lorentz factor are sufficiently large, following Eqs.~\ref{eq:xiB} and~\ref{eq:Gmin}. 

At large radii the external medium affects fireball evolution, and a "reverse shock" is driven backward into the fireball ejecta and decelerates it. For typical GRB fireball parameters this shock is also mildly relativistic \citep[e.g.][]{fireballs3}, and its parameters are similar to those of an internal shock with $\Delta t\sim10$~s. Protons may therefore be accelerated to $>10^{20}$~eV not only in the internal wind shocks, but also in the reverse shock \citep{WnB-AG,W01rev}. 
This implies that proton acceleration to $>10^{20}$~eV is possible, provided the constraints of Eqs.~\ref{eq:xiB} and~\ref{eq:Gmin} (with $\Delta t\sim10$~s) are satisfied, also in (the currently less favorable) scenario where GRB $\gamma$-rays are produced in the shock driven by the fireball into the surrounding gas, rather than by internal collisions \citep[as suggested, e.g., in][]{Dermer99}. 

The constraints given by Eqs.~(\ref{eq:xiB}) and (~\ref{eq:Gmin}) are
remarkably similar to those inferred from $\gamma$-ray
observations, based on independent physical arguments: $\Gamma>300$ is implied by the $\gamma$-ray spectrum by the requirement to avoid high pair-production optical depth, and magnetic field close to equipartition, $u_B/u_e\sim0.1$, is required in order to account for the observed $\gamma$-ray emission \citep{fireballs1,fireballs2,fireballs3}. This was the basis for the association of GRB's and UHECR's  suggested in \citep{W95a}. In the following sub-section we discuss the new constraints on model parameters implied by afterglow observations, and their implications.

\subsection{Implications of afterglow observations}
\label{sec:model_observations}

Afterglow observations lead to the confirmation of the
cosmological origin of GRBs and confirmed standard model
predictions of afterglow that results from synchrotron emission of
electrons accelerated to high energy in the highly relativistic
shock driven by the fireball into its surrounding gas. Afterglow observations provide therefore strong support for the  underlying fireball scenario. In addition, afterglow observations provide important information on the values of model parameters that enter the constraints given by Eqs.~(\ref{eq:xiB}) and (\ref{eq:Gmin}).

Prior to the detection of afterglows, it was commonly assumed that
the farthest observed GRB's lie at redshift $z\sim1$ \citep{MnP92,Piran92}. Based on afterglow redshift determinations, we now know that detected GRB's typically lie at farther distances \citep[e.g.][]{Bloom03}. This implies that the
characteristic GRB luminosity is higher by an order of magnitude compared to pre-afterglow estimates, $L_\gamma\approx10^{52}{\rm erg/s}$ instead of $L_\gamma\approx10^{51}{\rm erg/s}$. This relaxes the constraint on magnetic field energy fraction given by Eq.~(\ref{eq:xiB}). The implications of the revised GRB redshift distribution to the inferred GRB energy production rate are discussed in detail in \S~\ref{sec:rate}.

In several cases, fast follow up afterglow observations
allowed the detection of radio and optical emission from the reverse shock
\citep[][and references therein]{ZKM03,SR03}. These observations provide direct information on the plasma conditions in the reverse shock, where acceleration of protons
to high energy may take place (see \S~\ref{sec:model_description}). Two major conclusions were drawn from the analysis of the early optical and radio reverse shock emission. First, lower limits to the initial fireball Lorentz factors were inferred, in the range of $\Gamma>100$ to $\Gamma>1000$ \citep{ZKM03,SR03}. Second, the magnetic field in the reverse shock was inferred to be close to equipartition, that is $u_B/u_e$ was inferred to be of order unity \citep{Draine00,ZKM03}. Early afterglow observations provide therefore constraints on $\Gamma$ and on $u_B/u_e$ which are (i) Independent of the constraints derived from $\gamma$-ray observations; (ii) Consistent with the $\gamma$-ray constraints; and (iii) Are remarkably similar to the constraints of Eqs.~(\ref{eq:xiB}) and (\ref{eq:Gmin}), that need to be satisfied in order to allow proton acceleration to $>10^{20}$~eV.

\subsection{Comparison with other authors}
\label{sec:model_other}

\citet{Gallant99} and, more recently, \citet{Achterberg01} have
considered particle acceleration by the ultra-relativistic,
$\Gamma\sim300$ shock driven by the fireball into its surrounding
medium. They argue that protons can not be accelerated in this
{\it external ultra-relativistic} shock to ultra-high energy.
Regardless of whether or not this claim is valid, it is irrelevant
for the model proposed in \citep{W95a} and discussed in
\S~\ref{sec:model_description}, where proton acceleration takes
place in {\it internal (reverse) mildly-relativistic} shocks.
Similarly, the claims in \citet{Ostrowski02}, that the spectrum of
protons accelerated in {\it ultra-relativistic} shocks is much
steeper than $dn_p/dE_p\propto E_p^{-2}$, the spectrum expected
for sub-relativistic shocks and required to account for the
observed UHECR spectrum (see \S~\ref{sec:rate_CR}), are not
applicable to the model proposed in \citep{W95a}.

It should be emphasized here, that non-relativistic collisionless
shocks are observed in many types of astrophysical systems, and
that the theoretical understanding of particle acceleration in
such shocks \citep{Drury83,Blandford87} is more developed than in
the case of relativistic collisionless shocks. The GRB model for
particle acceleration, described in
\S~\ref{sec:model_description}, relies on our understanding of
acceleration in non-relativistic collisionless shocks, and
therefore is not subject to the uncertainties described in the
preceding paragraph, which are related to acceleration in
ultra-relativistic shocks.

Association of GRB's and UHECR's has been also suggested by Vietri
(1995) and by Milgrom \& Usov (1995), who noted that GRB's may
accelerate protons to $>10^{20}$~eV energy. While Milgrom \& Usov
did not suggest an acceleration mechanism, the criticism of
Gallant et al. (1999) may be relevant to the mechanism proposed by
Vietri. Their main point is that the fractional energy gain per
shock crossing is of order unity for highly relativistic shocks,
rather than of order $\Gamma^2$ as suggested in \cite{Vietri95}.
This implies that the acceleration time in ultra-relativistic
shocks is much longer than estimated by \cite{Vietri95}. However,
this conclusion does not necessarily imply that the acceleration
process is ineffective. In their estimate of the acceleration time
\cite{Gallant99} and \citet{Achterberg01} used an up-stream magnetic field
amplitude of $1\mu$G, typical to the inter-stellar medium. GRB
observations imply that this pre-shock magnetic field must be
amplified by many orders of magnitude in the GRB shock
\citep{Gruzinov99}, and it is therefore far from clear that the
magnetic field value relevant for particle deflection up-stream is
the un-perturbed pre-shock field. The up-stream magnetic field may
be amplified ahead of the shock by, e.g., the streaming of high
energy particles \citep{Bell01,Dermer02}, in which case
acceleration to ultra-high energy is possible.

\section{UHECR energy generation rate and spectrum}
\label{sec:rate}

\subsection{The GRB energy generation rate}
\label{sec:rate_GRB}

The GRB model for UHE cosmic-ray production was suggested prior to the detection of afterglows. Estimates of the rate of GRB's were based at that time on the $\gamma$-ray flux distribution, and ranged from $\sim3/{\rm Gpc^{3}yr}$ \citep{Piran92} to $\sim30/{\rm Gpc^{3}yr}$ \citep{MnP92}.
The estimated average $\gamma$-ray energy release in a single GRB, based on a characteristic peak flux of $\sim10^{51}{\rm erg/s}$ \citep[e.g.][]{Fishman95}, was $\sim10^{52}$~erg. These estimates were subject to large uncertainties, since the $\gamma$-ray luminosity function as well as the evolution of GRB rate with redshift were poorly constrained. Based on the rate and energy estimates, the rate of $\gamma$-ray energy generation by GRB's was estimated to be
$\sim10^{44}{\rm erg/Mpc^3yr}$. The determination of GRB redshifts, which was made possible by the detection of afterglows, allows a more reliable estimate. 

Most of the GRB's are observed from $z>1$, since they can be detected out to large redshift. This implies that the GRB rate density at $z>1$ is better constrained by the observations than the local, $z=0$, rate. The inferred local rate depends on the assumed redshift evolution. It is now commonly believed that the GRB rate evolves with redshift following the star-formation rate, based on the association of GRB's with type Ib/c supernovae. This association, which was originally motivated by the temporal and angular coincidence of GRB980425 and SN1998bw \citep{Galama98}, has gained significant support from the identification of a SN1998bw-like spectrum in the optical afterglow of GRB030329 \citep{Stanek03,Hjorth03}. It is also supported by evidence for optical supernovae emission in several GRB afterglows \citep{Bloom03a}. Adopting the assumption, that the GRB rate follows the redshift evolution of the star formation rate, the local ($z=0$) GRB rate density was inferred by \citet{Schmidt01} to be $R_{\rm GRB}(z=0)\approx 0.5\times{\,10^{-9}\rm Mpc^{-3}~yr^{-1}}$. A similar result was later obtained, under similar assumptions, by \citet{PSF03}. 

The local rate density determined by \citet{Schmidt01} and by \citet{PSF03} is uncertain due to uncertainties in the determination of the evolution of the star formation rate: A faster evolution (rate increase) with redshift implies a lower local, $z=0$, GRB rate. In their analysis, both \citet{Schmidt01} and \citet{PSF03} have not used, however, the detailed information provided by BATSE on the observed distribution of GRB peak fluxes, and the detailed shape of the observed GRB redshifts distribution \citep[][has used only the value of $<V/V_{\rm max}>$]{Schmidt01}. In a more detailed analysis, taking into account these constraints, \citet{GPW03} have shown that observations allow to discriminate between different assumptions regarding the evolution with redshift of the GRB rate density. Redshift evolution following the Rowan-Robinson \citep{RR99} star formation rate evolution 
($R_{\rm GRB}\propto 10^{0.75z}$ up to $z=1$) was found consistent with observations, while a more rapid evolution \citep[as assumed, e.g. in][]{PM01}, was found inconsistent. For the Rowan-Robinson evolution, \citet{GPW03} find $R_{\rm GRB}(z=0)\approx 0.5\times{\,10^{-9}\rm Mpc^{-3}~yr^{-1}}$. Given the current (systematic uncertainties in the redshift) data, this rate is accurate to within a factor of a few \citep{GPW03}.

The local energy generation rate in $\gamma$-rays by GRB's, $\dot{\varepsilon}_{\gamma}$, is given by the product of $R_{\rm GRB}(z=0)$ and the average $\gamma$-ray energy release in a single GRB, $\varepsilon_\gamma$. \citet{Bloom03} provide $\varepsilon_\gamma$ for 27 bursts with known redshifts, in a standard rest-frame bandpass, 0.02~MeV to 2~MeV. The average is $\varepsilon_\gamma=2.9\times10^{53}$~erg, with estimated uncertainty, due to the correction to a fixed rest-frame bandpass, of $\sim20\%$ for individual bursts (and much smaller for the average). In calculating $\dot{\varepsilon}_{\gamma}$ from this value of $\varepsilon_\gamma$, the following point should be taken into account. $\varepsilon_\gamma$ is the average energy for bursts with known redshift, most of which were localized by the BeppoSAX satellite. Since BeppoSAX has a higher detection flux threshold than BATSE \citep[see][]{Band03,GPW03}, it is sensitive to $\approx70\%$ of the bursts detectable by BATSE, for which the GRB rate $R_{\rm GRB}(z=0)$ was inferred. Thus, the energy generation rate by bursts detectable by BeppoSAX is 
\begin{equation}
\dot{\varepsilon}_{\gamma[0.02\rm MeV,2MeV]}^{\rm GRB}\approx 0.7R_{\rm GRB}(z=0)\varepsilon_\gamma= 10^{44} {\rm erg~Mpc^{-3}~yr^{-1}}. \label{eq:GRB_g_rate}
\end{equation}
As mentioned above, the main uncertainty in determining $\dot{\varepsilon}_{\gamma}$ is related to the uncertainty in the local GRB rate, due to which the value given in Eq.~\ref{eq:GRB_g_rate} is accurate to within a factor of few.

The energy generation rate given by Eq.~\ref{eq:GRB_g_rate} does not reflect the total energy emitted by GRB's in $\gamma$-rays, rather the energy emitted in the $[0.02\rm MeV,2MeV]$ band. Emission at higher energy has been detected in many GRB's \citep[see, e.g.][]{GRB941017}, implying that the total $\gamma$-ray energy emission is higher than that limited to the $[0.02\rm MeV,2MeV]$ band. A bandpass independent results may be obtained as follows. GRB $\gamma$-ray spectra are well described by broken power-laws, $dn_\gamma/dE_\gamma\propto E_\gamma^{-\alpha}$ for $E_\gamma<E_b$ and $E_\gamma^{-\beta}$ for $E_\gamma>E_b$ \citep{Band93,Preece98}. The observed (redshifted) break energy $E_b$ is typically a few hundered keV, $\beta\approx2$ and $\alpha\approx1$. The observed $\gamma$-rays are produced in the fireball model by synchrotron and inverse-Compton emission of electrons accelerated to high energy by collisionless shocks in the expanding fireball wind (see \S~\ref{sec:model} for more detail). The high energy part of the GRB spectrum is produced by energy loss of relativistic electrons, accelerated to a power-law distribution, $dn_e/dE_e\propto E_e^{-2}$. Taking into account the fact that the photon energy is proportional to the square of the electron energy (and hence that $\log(E_\gamma)$ spans twice the range spanned by $\log(E_e)$), and that the rest frame break energy is higher than observed by a factor $1+z$, the rate of  energy generation in relativistic electrons is
\begin{equation}
E_e^2\frac{d\dot{n}_e^{\rm GRB}}{dE_e}\approx\dot{\varepsilon}_{\gamma[0.02\rm MeV,2MeV]}^{\rm GRB} \approx10^{44}{\rm erg~Mpc^{-3}~yr^{-1}}. \label{eq:GRB_e_rate}
\end{equation} 

Finally, the following point should be mentioned. The numbers quoted above for the GRB rate density and $\gamma$-ray energy release are based on the assumption that GRB $\gamma$-ray emission is isotropic. If, as now commonly believed \citep[see][and references therin]{Frail01}, the emission is confined to a solid angle $\Delta\Omega<4\pi$, then the GRB rate is increased by a factor $(\Delta\Omega/4\pi)^{-1}$ and the GRB energy is decreased by the same factor. However, their product, the energy generation rate, is independent of the solid angle of emission.

\subsection{Comparison with UHECR observations}
\label{sec:rate_CR}

The cosmic-ray spectrum flattens at $\sim10^{19}$~eV
\citep{fly,agasa}. There are indications that the spectral change
is correlated with a change in composition, from heavy to light
nuclei~\citep{fly,composition,HiResMIA}. These characteristics,
which are supported by analysis of Fly's Eye, AGASA and HiRes-MIA
data, and for which some evidence existed in previous
experiments~\citep{Watson91}, suggest that the cosmic ray flux is
dominated at energies $< 10^{19}$~eV by a Galactic component of
heavy nuclei, and at UHE by an extra-Galactic source of protons.
Also, both the AGASA and Fly's Eye experiments report an
enhancement of the cosmic-ray flux near the Galactic disk at
energies $\le10^{18.5}$~eV,  but not at higher
energies~\citep{anisotropy,agasa1}. Fly's Eye stereo spectrum is
well fitted in the energy range $10^{17.6}$~eV to $10^{19.6}$~eV
by a sum of two power laws: A steeper component, with differential
number spectrum $J\propto E^{-3.50}$, dominating at lower energy,
and a shallower component, $J\propto E^{-2.61}$, dominating at
higher energy, $E>10^{19}$~eV. The data are consistent with the
steeper component being composed of heavy nuclei primaries, and
the lighter one being composed of proton primaries.

\citet{BnW02} have shown that the observed UHECR flux
and spectrum may be accounted for by a two component, Galactic +
extra-Galactic model. For the Galactic component, this model
adopts the Fly's Eye fit,
\begin{equation}
\frac{dn}{dE} ~\propto~ E^{-3.50}. \label{eq:galacticspectrum}
\end{equation}
The spectrum and energy generation rate of extra-Galactic protons are given in this model by
\begin{equation}
E_p^2\frac{d\dot{n}_p^{\rm CR}}{dE_p}=0.65\times 10^{44} {\rm
erg~Mpc^{-3}~yr^{-1}}. \label{eq:GRB_p_rate}
\end{equation}
The spectral index, 2, is that expected for acceleration
in sub-relativistic collisionless shocks in general, and in
particular for the GRB model discussed in
\S~\ref{sec:model_description}. The energy generation rate integrated over the energy range of $10^{19}$~eV to $10^{21}$~eV is
\begin{equation}
\dot{\varepsilon}^{\rm CR}_{[\rm10^{19},10^{21}]eV}= 3\times 10^{44} {\rm
erg~Mpc^{-3}~yr^{-1}}. \label{eq:energyrate}
\end{equation}
Uncertainties in the absolute energy calibration of the experiments lead to uncertainty of $\approx20\%$ in this rate \citep{BnW02}.

The model used in \citep{BnW02} is similar to that proposed in \citet{W95b}. The improved constraints on UHECR spectrum and flux provided by the recent observations of HiRes do not change the estimates given in \citep{W95b} for the energy generation rate and spectrum, Eq.~(\ref{eq:GRB_p_rate}), but  reduce the uncertainties. Comparing Eqs.~(\ref{eq:GRB_p_rate}), or~(\ref{eq:energyrate}), and (\ref{eq:GRB_e_rate}) (which, as explained in \S~\ref{sec:rate_GRB}, is accurate to within a factor of a few) we find that the rate at which GRB's produce energy in accelerated high energy electrons is comparable to the rate at which energy should be produced in high energy protons in order to account for the observed UHECR flux.

It is important to emphasize the following point. As explained in \S~\ref{sec:rate_GRB}, the $\gamma$-ray energy production rate reflects the rate at which energy is produced by GRB's in relativistic electrons. The ratio between the energy carried by relativistic electrons and protons is not known from basic principles \citep[observations suggest that in sub-relativistic shocks protons carry $\sim10$ times more energy than electrons, e.g.][]{Blandford87,SN1006CR}. Thus, an exact match between the $\gamma$-ray and UHECR generation rates should not in general be expected. Rather, the two rates are expected to be similar to within an order of magnitude.

\subsection{Comparison with other authors}
\label{sec:rate_other}

Recent claims, that the GRB energy generation rate is too
small to account for the observed UHE cosmic-ray flux,
\citep{Stecker00,Berezinsky02}, are not valid.

In \citep{Stecker00} it is argued that at most "10\% of the cosmic
rays observed above $10^{20}$~eV can be accounted for by GRB's."
The UHECR energy generation rate estimated in \citep{Stecker00} is
similar to the rate given by Eq.~(\ref{eq:GRB_p_rate}), as derived
in \citep{W95b,BnW02}, but the GRB $\gamma$-ray energy generation
rate estimated in \citep{Stecker00} is smaller by approximately an
order of magnitude compared to the rate derived here,
Eq.~(\ref{eq:GRB_g_rate}). The $\gamma$-ray energy generation rate
by GRB's is underestimated in \citep{Stecker00} by an order of
magnitude, since it is based on an earlier, less accurate estimate of the GRB
rate per unit volume, compare \citep{Schmidt99} to \citep{Schmidt01,Frail01,GPW03}, 
and neglects GRB $\gamma$-rays in the 1~MeV to 2~MeV band.

Berezinsky (2002) claims that the required UHECR generation rate
is three orders of magnitude larger than the GRB $\gamma$-ray
energy generation rate. He argues that the energy generation rate
of UHECR's implied by observations is $\approx 3\times 10^{46}
{\rm erg~Mpc^{-3}~yr^{-1}}$. The discrepancy with our derived
rate, Eq.~(\ref{eq:energyrate}), arises because Berezinsky assumes
that extra-Galactic protons dominate the observed flux in the
range of $10^{17}$~eV to $10^{18}$~eV. However, most observers
attribute the flux in this energy range to Galactic cosmic rays.
Thus, in our model,
Eqs.~(\ref{eq:galacticspectrum}--\ref{eq:GRB_p_rate}),
extra-Galactic sources dominate the flux only above $10^{19}$~eV,
and the fraction of the observed cosmic-ray flux that is
contributed by the extra-Galactic component at the energy range of
$10^{17}$~eV to $10^{18}$~eV is
\begin{equation}
\frac{J^{\rm ex-Galac.}}{J^{\rm Galac.}}\sim
10^{-2}\left(\frac{E}{10^{17}\rm eV}\right). \label{eq:fraction}
\end{equation}
Observations strongly suggest that cosmic-rays above and below
$10^{19}$~eV originate from different sources: Galactic sources of
heavy nuclei are likely to dominate below $10^{19}$~eV, while
extra-Galactic proton sources are likely to dominate at higher
energy. Thus, by assuming that extra-Galactic particles dominate
the flux down to $10^{17}$ eV, Berezinsky greatly overestimates
the required energy for extra-galactic sources. In addition, the
GRB energy generation rate is underestimated in
\citep{Berezinsky02} in the same way as in \citep{Stecker00}.

\section{Discussion.}
\label{sec:discussion}

The main constraints that a relativistic wind (fireball) need to satisfy to allow proton acceleration to $>10^{20}$~eV are given by Eqs.~(\ref{eq:xiB}) and (\ref{eq:Gmin}): The magnetic field energy density $u_B$ should exceed a few percent of the relativistic electron energy density $u_e$, and the wind Lorentz factor $\Gamma$ should exceed $\approx10^2$. The similarity of these constraints and the constraints imposed on wind parameters, based on independent physical considerations, by $\gamma$-ray observations were the basis for the association of GRB and UHECR sources suggested in \citep{W95a}. Afterglow observations have shown that the characteristic GRB luminosity is higher than estimated based on $\gamma$-ray observations alone, $10^{52}{\rm erg/s}$ instead of $10^{51}{\rm erg/s}$, relaxing the constraint of Eq.~(\ref{eq:xiB}) on $u_B/u_e$. In addition, early optical and radio afterglow observations provide new constraints on wind parameters: They imply large Lorentz factors, $\Gamma>10^2$ to $\Gamma>10^3$, and large magnetic field energy density in the fireball plasma, $u_B/u_e$ of order unity (see \S~\ref{sec:model_observations}). These constraints are consistent with those previously inferred from $\gamma$-ray observations, and with the constraints imposed by the requirement to allow proton acceleration to $>10^{20}$~eV.

The local, $z=0$, $\gamma$-ray energy generation rate by GRB's is $\approx10^{44}{\rm erg/Mpc^3yr}$ (see Eqs.~(\ref{eq:GRB_g_rate}), (\ref{eq:GRB_e_rate})). Afterglow observations allow to determine this rate to within a factor of a few. The main uncertainty is due to uncertainties in the redshift evolution of the GRB rate (see \S~\ref{sec:rate_GRB}). The local rate of energy generation in high energy protons, required to account for the observed UHECR flux, is determined by observations with smaller uncertainty, and is also $\approx10^{44}{\rm erg/Mpc^3yr}$ (see \S~\ref{sec:rate_CR}, eqs.~(\ref{eq:GRB_p_rate}), (\ref{eq:energyrate})). 

Prior to direct GRB redshift measurements, which became possible with the detection of afterglows, estimates of the GRB rate ranged from $\sim3/{\rm Gpc^{3}yr}$ \citep{Piran92} to $\sim30/{\rm Gpc^{3}yr}$ \citep{MnP92}. These estimates were subject to large uncertainties, since the $\gamma$-ray luminosity function as well as the evolution of GRB rate with redshift were poorly constrained. The measurements of GRB redshifts allow a more accurate estimate, $0.5/{\rm Gpc^{3}yr}$ at $z=0$. Although this rate is significantly lower than the pre-afterglow estimates, the estimated rate of energy generation in $\gamma$-rays by GRBs is similar to the pre-afterglow estimate, which was $\sim10^{44}{\rm erg/Mpc^3yr}$. This is due to the fact that the average $\gamma$-ray energy release in a single GRB is larger than the pre-afterglow estimates, $10^{53.5}$~erg instead of $\sim10^{52}$~erg. 

The numbers quoted above for the GRB rate density and $\gamma$-ray energy release are based on the assumption that GRB $\gamma$-ray emission is isotropic. If, as now commonly believed \citep[see][and references therin]{Frail01}, the emission is confined to a solid angle $\Delta\Omega<4\pi$, then the GRB rate is increased by a factor $(\Delta\Omega/4\pi)^{-1}$ and the GRB energy is decreased by the same factor. However, their product, the energy generation rate, is independent of the solid angle of emission. 

The local GRB rate implies that the rate of GRB's out to a distance from which most protons of energy exceeding $10^{20}$~eV originate, $\simeq90$~Mpc \citep[see figure 2 in][]{W95b}, is $\sim10^{-3}/$yr. The number of GRB's contributing to the observed flux at any given time is given by the product of this rate and the spread in arrival time of protons, due to the combined effect of stochastic propagation energy loss and deflection by magnetic fields \citep{W95a}. This time spread may be as large as $10^7$~yr for $10^{20}$~eV originating at $90$~Mpc distance, implying that the number of GRBs contributing to the $>10^{20}$~eV flux at any given time may reach $\sim10^4$ \citep{W01rev}. The upper limit on the strength of the inter-galactic magnetic field, combined with the low local rate of GRB's, leads to unique predictions of the GRB model for UHECR production
\citep{MnW96,WnM96}, that may be tested with operating
\citep{HiRes}, under-construction \citep{auger} and planned
\citep{TA} large area UHECR detectors. In particular, a critical
energy is predicted to exist, $10^{20}{\rm eV}\le
E_c<4\times10^{20}{\rm eV}$, above which a few sources produce
most of the UHECR flux, and the observed spectra of these sources
is predicted to be narrow, $\Delta E/E\sim1$: The bright sources
at high energy should be absent in UHECRs of much lower energy,
since particles take longer to arrive the lower their energy. The
model also predicts the emission of high-energy, $>1$~TeV
neutrinos \citep{WnB97}, a prediction that may be tested with
operating (AMANDA), under-construction (ANTARES,IceCube,NESTOR)
and planned (NEMO) large volume neutrino telescopes (see Halzen \&
Hooper 2002 for review). For more detailed discussion of model
predictions, see \citep{W01rev}.

At energies $> 10^{20}$~eV, the predicted number, $N$, of events
in conventional models is uncertain due to the unknown clustering
scale, $r_0$, of the sources, $\sigma(N_{\rm predicted})/N_{\rm
predicted} = 0.9(r_0/{10~\rm Mpc})^{0.9}$ ~\citep{clustering}. For
GRB's in particular, the flux above $3\times10^{20}$~eV is
expected to be dominated by few sources \citep{MnW96}, and hence
large deviations from a homogeneous source distribution may be
expected (see \citet{W01rev} for detailed discussion).

\paragraph*{Acknowledgments.}
EW thanks J. N. Bahcall for useful discussions, and the IAS for
hospitality. This research was partially supported by MINERVA and AEC grants.

\end{document}